# Microsecond Valley Lifetime of Defect-Bound Excitons in Monolayer WSe$_2$

Galan Moody[1,*], Kha Tran[2], Xiaobo Lu[3], Travis Autry[1], James M. Fraser[4], Richard P. Mirin[1], Li Yang[3], Xiaoqin Li[2], and Kevin L. Silverman[1]

[1] *National Institute of Standards & Technology, Boulder, CO 80305, USA.*
[2] *Department of Physics and Center for Complex Quantum Systems, University of Texas at Austin, Austin, TX 78712, USA.*
[3] *Department of Physics and Institute of Materials Science and Engineering, Washington University in St. Louis, St. Louis, MS 63136, USA.*
[4] *Department of Physics, Engineering Physics & Astronomy, Queen's University, Kingston, Ontario K7L 3N6, Canada.*
*corresponding author: galan.moody@nist.gov

In atomically thin two-dimensional semiconductors such as transition metal dichalcogenides (TMDs), controlling the density and type of defects promises to be an effective approach for engineering light-matter interactions. We demonstrate that electron-beam irradiation is a simple tool for selectively introducing defect-bound exciton states associated with chalcogen vacancies in TMDs. Our first-principles calculations and time-resolved spectroscopy measurements of monolayer WSe$_2$ reveal that these defect-bound excitons exhibit exceptional optical properties including a recombination lifetime approaching 200 ns and a valley lifetime longer than 1 μs. The ability to engineer the crystal lattice through electron irradiation provides a new approach for tailoring the optical response of TMDs for photonics, quantum optics, and valleytronics applications.

(Dated: 14 June 2018)

Solid-state quantum technologies rely on the ability to store and transmit information using the internal quantum degrees of freedom of electrons and excitons, such as spin or total angular momentum [1]. In two-dimensional transition metal dichalcogenide (TMD) crystals, information can be also encoded in the valley pseudospin associated with the occupancy of distinct valleys in momentum space [2]. In monolayer TMDs, broken inversion symmetry and strong spin-orbit interaction couple the electronic spin and valley pseudospin, leading to valley-dependent optical selection rules for the direct-gap transitions at the $K$ and $K'$ valleys. Due to spin-valley coupling, excitons residing in the $K$ ($K'$) valley interact with left-circularly (right-circularly) polarized light, enabling optical access to the valley pseudospin [3]. The ability to coherently control valley excitons in monolayer TMDs has revived interest in the field of valleytronics, with seminal examples including the optical generation of valley polarization [4,5] and valley coherence [6,7], optical rotations of the valley pseudospin [8–10], and the valley Hall effect of photo-excited carriers [11,12].

TMD-based valleytronics have typically relied on light-matter interactions with delocalized excitons in monolayer crystals. Excitons have an intrinsically fast recombination lifetime and dephasing time due to their large interband oscillator strength, which place strict limitations on the ability to coherently manipulate the valley pseudospin before information is lost through dissipation or decoherence [13]. Consequently, longer-lived excitonic transitions with stable valley pseudospin are required to make significant breakthroughs for practical quantum device applications. Promising candidates exhibiting long valley polarization lifetimes include interlayer excitons in heterostructures [14], optically dark excitons [15], and photo-excited valley-polarized carriers [16,17].

An interesting question is whether, like delocalized excitons in monolayers and heterostructures, spin-valley coupling is preserved for spatially localized excitons. Point defects naturally present in the TMD crystal lattice serve as zero-dimensional quantum dot sites that can confine excitons, extending the ~1 ps delocalized exciton recombination lifetime up to ~1 ns [18–22]. The longer lifetime suggests that the valley polarization lifetime may also be enhanced, although it is not clear *a priori* whether defects mix the $K$ and $K'$ valleys. Moreover, the lack of control over the presence and spatial position of defects in TMDs currently limits their utility for scalable devices. Addressing these challenges would facilitate the realization of devices with desirable optical and electronic properties for valleytronics and quantum technologies.

In this Letter, we demonstrate a simple and reliable method for engineering defects that inherit the spin-valley coupling of the host TMD crystal. Using an electron beam, we irradiate a pristine monolayer of WSe$_2$, which introduces an inhomogeneously broadened defect resonance in the emission spectrum below the delocalized exciton. Irradiation preferentially creates chalcogen vacancies [23] that give rise to multiple defect bands in the electronic bandstructure as verified with density-functional theory (DFT) calculations. Using time-resolved photoluminescence spectroscopy, we find that the recombination lifetime of defect-bound excitons is as long as 200 ns – at least two orders of magnitude longer than natural tungsten-based defects often present in monolayer WSe$_2$ [24,25]. Moreover, the defect emission exhibits large circular dichroism that persists for at least 1 μs, confirming *ab initio* calculations predicting stable valley polarization of defect states including exciton effects [26]. Because of the relatively simple processing steps, the few-nanometer spatial resolution of an electron beam, and the widespread availability of electron microscopes, defect engineering via electron irradiation is a promising approach for modifying the optical response of monolayer TMDs and may allow for the creation of site-controlled quantum dots as sources of single photons, entangled-photon pairs, and long-lived valley polarized excitons [21,27].

To characterize the effects of electron-beam irradiation on the optical response of monolayer $WSe_2$ (Fig. 1(a)), we first measure a baseline optical spectrum of a pristine flake, which was obtained through mechanical exfoliation onto an $SiO_2$/silicon substrate. The photoluminescence (PL) spectrum, taken at 5 K with 2.33 eV continuous-wave excitation and a ~1 μm spot size on the sample, is shown in Fig. 1(b) by the dashed curve. The two prominent peaks are assigned to neutral ($X^0$) and negatively charged ($X^-$) excitons [28], whereas the series of peaks at lower energy are associated with native defects in the material. Interestingly, low-temperature scanning tunneling microscopy reveals that these defect emission bands are associated with tungsten vacancies, as opposed to chalcogen vacancies that prevail in other types of TMDs [24].

The same flake was then uniformly irradiated with an electron beam in a scanning electron microscope with ~$10^7$ electrons/μm$^2$ (30 kV accelerating voltage, 350 pA current, ~10 nm spot size). Irradiation dramatically alters the PL spectrum: while the $X^0$ and $X^-$ resonance energies are unaffected, their relative quantum yield is reduced by a factor of ~5, and a new emission band (*D*) appears ~130 meV below the $X^0$ resonance with a ~100 meV inhomogeneous linewidth (solid line in Fig. 1(b)). Identical spectra were obtained from multiple irradiated flakes from different substrates with no noticeable change to the spectrum after storing the samples under ambient conditions for at least three months. Exposure to electron irradiation has been shown experimentally and theoretically to create defects in TMDs resulting from the ballistic displacement of atoms—favoring chalcogen vacancy production due to the larger cross section (lower atomic mass) of selenium and sulfur compared to the transition metals [23,29,30]. The sublinear increase of the intensity of peak *D* with excitation power ($\propto P^{0.7}$), compared to the linear scaling of peaks $X^0$ and $X^-$ shown in Fig. 1(c), suggests that this emission arises from defect-bound excitons localized near the selenium vacancy sites [31,32]. Similar changes to the optical response occur after electron-beam lithography, alpha-particle irradiation, and argon plasma treatment of TMDs [33–36]; however, the recombination dynamics and spin-valley coupling of engineered defects in TMDs have not been investigated, which is the focus of this Letter.

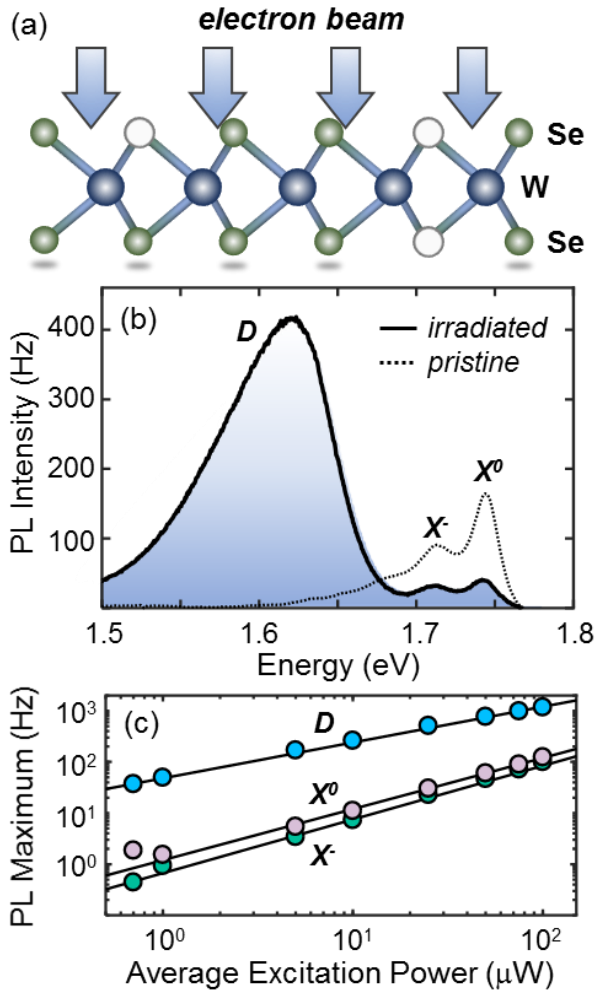

FIG. 1. (a) Electron-beam irradiation of monolayer $WSe_2$ creates defects associated with selenium vacancies in the crystal lattice. (b) Photoluminescence spectrum acquired for as-exfoliated (pristine, dashed line) $WSe_2$ at 5 K. The two prominent peaks correspond to the neutral ($X^0$) and negatively charged ($X^-$) excitons. Following electron irradiation, a broad spectral band (*D*) appears at lower energy due to emission associated with selenium vacancies. (c) Scaling of the maximum intensity of the neutral, charged, and defect exciton peaks with average power of the continuous-wave excitation. The $X^0$ and $X^-$ peaks scale linearly with exponent $k = 1$. The defect peak *D* scales sub-linearly with exponent $k = 0.7$.

First, we examine how the presence of selenium defects modifies the host crystal band structure and absorption spectrum through calculations within the DFT framework [29,37]. The first-principles calculations are performed by the Vienna Ab initio simulation package (VASP) with the Perdew, Burke, and Ernzerhof (PBE) functional [38,39]. The ground-state wave functions and eigenvalues are obtained by solving the Kohn-Sham equation with projector augmented wave (PAW) potentials. A 6-by-6 supercell is generated to mimic isolated defects. The plane-wave basis is set with a cutoff energy of 400 eV with a 2-by-2-by-1 $k$-point grid in the reciprocal space for converged DFT results. A vacuum space between neighboring layers is set to be more than 16 Å to avoid artificial interactions between neighboring slab structures.

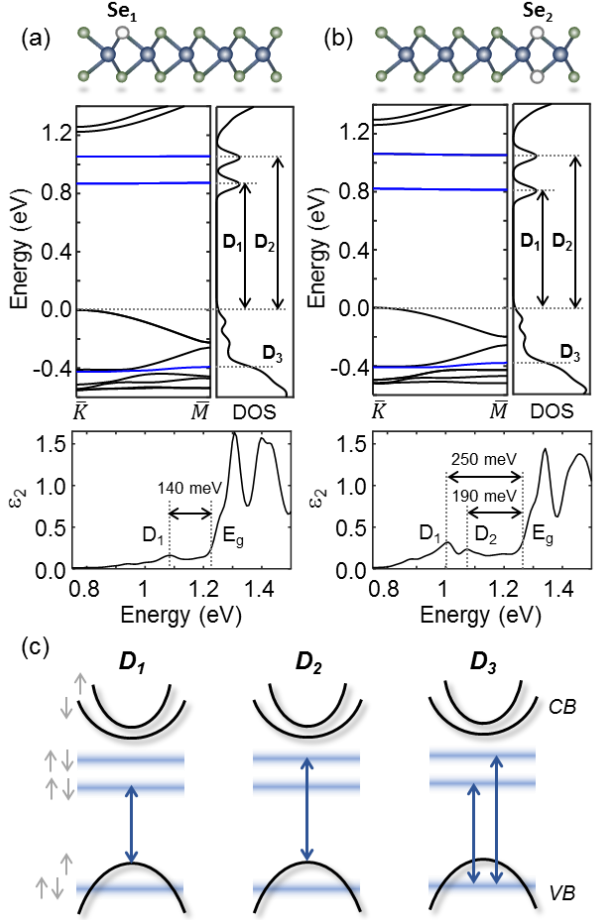

FIG. 2. DFT calculations of the electronic band structure, projected density of states, and linear absorption spectrum ($\varepsilon_2$) for the 6-by-6 supercell Brillouin zone for (a) a single selenium vacancy (Se$_1$) and (b) a double selenium vacancy (Se$_2$). The $K$ point of the original Brillouin zone is folded back to the $\bar{K}$ point of the 6-by-6 supercell. The blue curves in the band diagram are attributed to the to the defects. (c) Schematic of the band-to-band transitions near the $K$ valley that predominantly contribute to the emission spectrum.

Although TMDs can feature a rich variety of defects, we inspect the two most stable types in WSe$_2$ with the lowest enthalpy of formation—a single selenium vacancy (Se$_1$, Fig. 2(a)) and a double selenium vacancy (Se$_2$, Fig. 2(b)) [24]. The DFT-calculated bandstructure along the $\bar{K}$-to-$\bar{M}$ direction in the 6-by-6 supercell Brillouin zone is shown in the middle panels of Figs. 2(a) and 2(b) for Se$_1$ and Se$_2$, respectively. We find that the presence of selenium vacancies results in two additional unoccupied, mid-gap bands (D$_1$ and D$_2$) and at least one occupied band below the valence band maximum (D$_3$). Each defect band comprises two quasi-degenerate electronic spin-up and spin-down states. Electrons and holes occupying the defect bands are localized around the defect sites and introduce additional features below the bandgap ($E_g$), as illustrated in the absorption spectrum in the bottom panels of Figs. 2(a) and 2(b). In the absorption spectrum, peaks D$_1$ and D$_2$ are associated with transitions between the valence band maximum and each unoccupied defect band near the $K$ and $K'$ valleys, as illustrated in Fig. 2(c). The absorbance at energies above peaks D$_1$/D$_2$ and below the bandgap $E_g$ likely arises from the multiple D$_3$-type transitions between occupied and unoccupied defect bands [26]. Recombination at different energies associated with each defect type in Fig. 2(c) from both Se$_1$ and Se$_2$ at least partially contributes to the inhomogeneous broadening of the PL emission spectrum in Fig. 1.

While DFT calculations typically underestimate the band gap and do not include excitonic effects, higher-order GW approximations for the quasiparticle energies and the Bethe-Salpeter Equation (BSE) may change the defect levels. According to a recent study [26], these many-electron effects will modify the energy difference of the optical transitions between the conduction band edge and the unoccupied defect bands by ~15%. As a result, including many-electron effects will moderately change the optical spectra but will not qualitatively change the physical picture [29]. When including excitonic effects in the BSE framework, the binding energy and Bohr radius of delocalized and defect-bound excitons are nearly identical [24]. Moreover, the similarity between the defect transition energies and the quasiparticle bandgap is expected to strongly hybridize some defect-bound excitons with delocalized excitons in the pristine crystal lattice [26]. As a result, each excitonic state can have contributions from multiple bands, which impacts the spin-valley coupling and recombination dynamics as shown below.

Signatures of these different defect transitions are observed in time-resolved and polarization-resolved PL spectra. Time-resolved photoluminescence (TRPL) spectra are acquired using a 532-nm pulsed excitation source from a 100 fs, Ti:sapphire-pumped optical parametric oscillator with ~1 μJ/cm$^2$ fluence at the sample. An acousto-optical deflector is used to select every 32$^{nd}$ pulse (~420 ns period between excitation pulses) to avoid build-up of carriers in the defect states through re-excitation from subsequent pulses. The dynamics are recorded using a silicon single-photon avalanche diode and time-tagging electronics that are placed after a monochromator filter with ~1 nm bandwidth. The instrument resolution (full-width at half-maximum) is ~180 ps. The TRPL dynamics are shown in Fig. 3 for three representative emission energies across the defect band. At energies below ~1700 meV, the recombination dynamics exhibit a biexponential decay. Parameters from the fits to the data are shown in Fig. 3(b). As the emission energy decreases, both the fast ($T_F$) and slow ($T_S$) lifetimes increase. The lowest energy defect transitions near 1550 meV have a dominant slow component that decays in ~200 ns. With increasing emission energy, the relative amplitude of the fast component increases (Figure 3b), resulting in the highest-energy defect transition recombining in < 1 ns. The biexponential decay may be attributed to emission arising from thermalized optically bright and dark transitions or Auger-type processes that give rise to similar dynamics for neutral delocalized excitons, although further studies are needed to determine the mechanisms responsible.

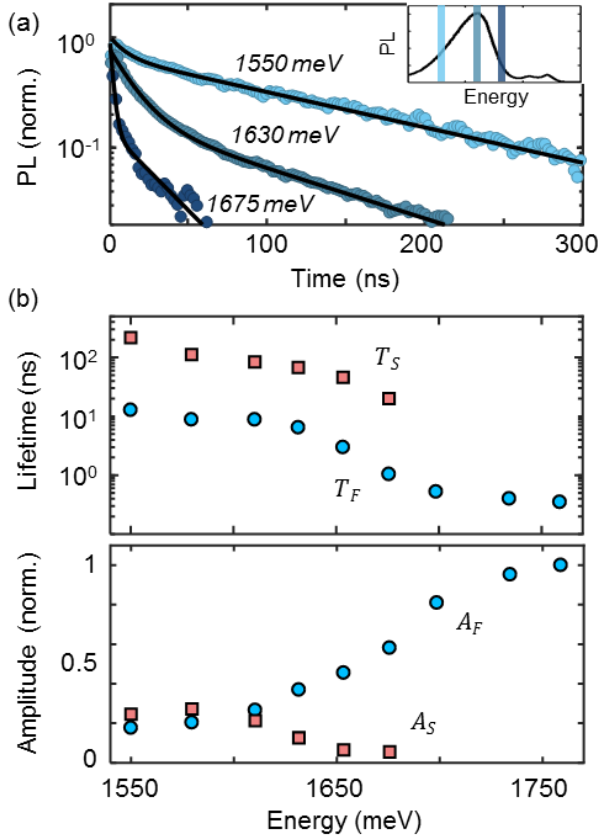

FIG. 3. (a) Time-resolved photoluminescence dynamics for three different emission energies within the defect band (points). The curves are fit with a biexponential decay function (solid lines). (b) Lifetimes and amplitudes of the slow (squares) and fast (circles) decay components from the biexponential fits.

Next, we examine the polarization selection rules of the defect resonance. Optically exciting the sample with left-circularly polarized light at 532 nm, we measure both the left- (+) and right-circularly polarized (−) emission, shown in Fig. 4(a). The defect emission exhibits a large degree of circular polarization, defined as $P_c = (I_{++} - I_{+-})/(I_{++} + I_{+-})$, where $I_{++}$ ($I_{+-}$) corresponds to co-circularly (cross-circularly) polarized excitation and detection. Interestingly, $P_c$ is highly sensitive to the emission energy, with maximum polarization up to 30% between 1550 meV and 1650 meV and nearly zero polarization for defect states emitting at higher energies. A large polarization is consistent with the valley-selective circular dichroism predicted from *ab initio* calculations [26]. With recombination lifetimes approaching 200 ns for defects emitting at 1580 meV, the 30% degree of circular polarization indicates that these states are at least partially protected from inter-valley scattering. In a simple picture, the steady-state polarization can be defined as $P_c = P_0/(1 + \tau/\tau_s)$, where $P_0$ is the initial degree of circular polarization immediately following optical excitation, $\tau$ is the recombination lifetime, and $\tau_s$ is the inter-valley scattering time [40]. By time-resolving the circularly polarized emission, shown in Fig. 4(b), we find that $P_0 = 34\%$ and is nearly constant during recombination. Using this initial polarization value for $P_0$, the time-integrated value of $P_c = 30\%$ from the steady-state spectrum, and $\tau = 200$ ns, we can place a lower bound on the valley lifetime of $\tau_s \geq 1$ μs.

The emission energy dependence of the valley polarization and its long lifetime for low-energy defects are consistent with the predictions of spin-valley coupling for defect-bound excitons [26]. Emission on the low-energy side of the resonance primarily involves $D_1$ and $D_2$ type defects with optical selection rules determined by transitions at the K and K' valleys. In the absence of inter-valley scattering, these transitions are expected to retain spin-valley coupling and exhibit large circular dichroism [26], consistent with the polarization-resolved spectra in Fig. 4. Emission at higher energy, however, is likely associated with $D_3$-type transitions between occupied and unoccupied defect bands. Because these defect states lack the intrinsic three-fold rotational symmetry of the host crystal, they extend throughout *k*-space and do not exhibit any spin-valley coupling, which is consistent with the absence of circularly polarized emission at higher energies. The presence of multiple types of defect transitions may also explain the emission energy dependence of the recombination dynamics shown in Fig. 3. Due to the similarity between the energies of the $D_3$-type defect and delocalized excitons, hybridization leads to additional oscillator strength for these transitions [26]. As a result, these defects exhibit a faster recombination lifetime than the $D_1$- and $D_2$-type defects emitting at lower energy.

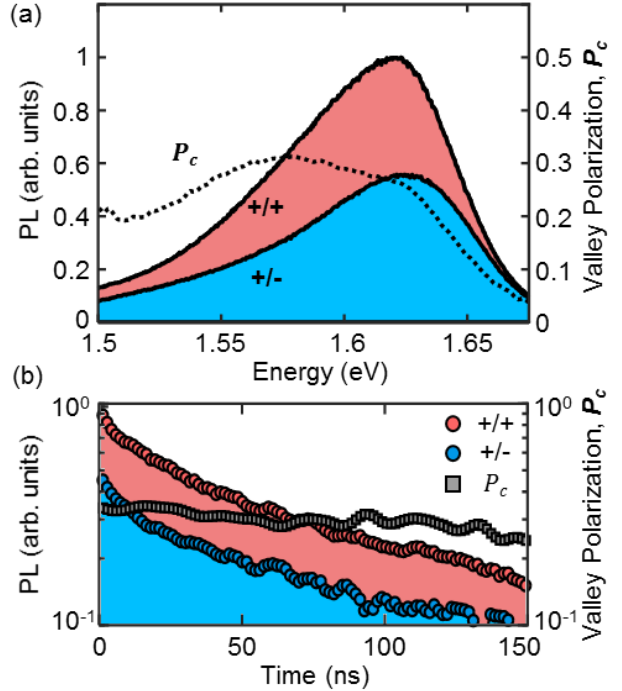

FIG. 4. (a) A comparison of the co-circularly (+/+) and cross-circularly (+/-) polarized photoluminescence spectrum (solid lines) reveals a maximum valley polarization of $P_c = 30\%$ (dashed line). (b) Normalized co-circularly (red circles) and cross-circularly (blue circles) polarized dynamics and the degree of circularly polarization $P_c$ (squares).

In summary, we have demonstrated that electron-beam irradiation is a promising approach for engineering defects in monolayer $WSe_2$. These defects arise from selenium vacancies as opposed to tungsten vacancies naturally present in the material. Excitons bound to selenium-based defects in $WSe_2$ exhibit ultralong recombination lifetimes (200 ns) and robust valley polarization with an intervalley scattering time of at least 1 μs—comparable to or longer than other promising approaches being pursued with bright (~40 ns) and dark (~1 μs) excitons in TMD heterostructures [14,15]. The intrinsically high nanometer spatial resolution of an electron beam could facilitate the fabrication of site-controlled single-quantum emitters in TMDs for quantum optics and photonics applications. These defect states in $WSe_2$ may have superior optical properties compared to previously studied defects and single emitters. For example, using the fast lifetime component of $\tau$ ~10 ns as a conservative estimate, a transform-limited coherence time of $2\tau = 20$ ns (corresponding to < 0.04 μeV or < 10 MHz homogeneous linewidth) is possible. Isolating single quantum emitters of this type through strain engineering would enable a scalable production of single-photon emitter arrays. Resonant nonlinear optical spectroscopy of these states might also reveal novel spin and valley dynamics, couplings, and multi-exciton transitions that can be leveraged for TMD-based quantum devices.

K. Tran acknowledges partial support by the National Science Foundation through the Center for Dynamics and Control of Materials: an NSF MRSEC under Cooperative Agreement No. DMR-1720595. X. Li also acknowledges support from the Welch Foundation F-1662. X. Lu and L. Yang acknowledge the support from the National Science Foundation (NSF) CAREER Grant No. DMR-1455346 and the Air Force Office of Scientific Research (AFOSR) grant No. FA9550-17-1-0304. This work is an official contribution of NIST, which is not subject to copyright in the United States.